\documentclass[a4paper,numbib, final, twoside, titelpage]{imaiai}%%%%where imaiai is the template name

\usepackage{graphicx} %for dealing with figures.
\usepackage{subfig} % for getting the subfigures e.g., "Figure 1a 

\usepackage{xcolor}
\usepackage{hyperref}
\hypersetup{
colorlinks,
linkcolor={black},
citecolor={blue!50!black},
urlcolor={blue!80!black},
}

\begin{document}

\title{Modular decomposition of protein structure using community detection}
\shorttitle{Modular decomposition of protein structure }
\shortauthorlist{W.P. Grant and S. E. Ahnert} %%% for verso running head

% \title[Modular decomposition of protein structure]{Modular decomposition of protein structure using community detection}

% \number{0000}

\author{{%%%% First author details
\sc William P. Grant}
% Need authors name and affiliation
$^\dagger$,\\[2pt]
Theory of Condensed Matter Group, Cavendish Laboratory, University of Cambridge, Cambridge, CB3 0HE.\\
$^\dagger${\email{Corresponding author: wpg23@cam.ac.uk}}\\[10pt]
%%%%%%/% Second author details
% \and\\
{\sc Sebastian E. Ahnert}\\[2pt]
Theory of Condensed Matter Group, Cavendish Laboratory, University of Cambridge, Cambridge, CB3 0HE and\\
Sainsbury Laboratory, University of Cambridge, Cambridge CB2 1LR}

% \author[W. P. Grant and S. E. Ahnert]{William P. {Grant}\affil{\dag}\\
% \add{Theory of Condensed Matter Group, Cavendish Laboratory, University of Cambridge,\\ Cambridge CB3 0HE, UK}\\
% \corr{\dag}{wpg23@cam.ac.uk}\\
% \nd\\
% Sebastian E. {Ahnert}\\
% \add{Theory of Condensed Matter Group, Cavendish Laboratory, University of Cambridge, Cambridge CB3 0HE, UK and Sainsbury Laboratory, University of Cambridge, Cambridge CB2 1LR, UK}\\
% \editedby{Ernesto Estrada}}
% [Received on 4 March 2018; editorial decision on 29 June 2018; accepted on 4 July 2018]
% \received{4 March 2018}
% \nullofn

\maketitle

\begin{abstract}
{As the number of solved protein structures increases, the opportunities for meta-analysis of this dataset increase too. Protein structures are known to be formed of domains; structural and functional subunits that are often repeated across sets of proteins. These domains generally form compact, globular regions, and are therefore often easily identifiable by inspection, yet the problem of automatically fragmenting the protein into these compact substructures remains computationally challenging. Existing domain classification methods focus on finding subregions of protein structure that are conserved, rather than finding a decomposition which spans the full protein structure. However, such a decomposition would find ready application in coarse-graining molecular dynamics, analysing the protein's topology, in \textit{de novo} protein design and in fitting electron microscopy maps. Here, we present a tool for performing this modular decomposition using the Infomap community detection algorithm. The protein structure is abstracted into a network in which its amino acids are the nodes, and where the edges are generated using a simple proximity test. Infomap can then be used to identify highly intra-connected regions of the protein. We perform this decomposition systematically across 4000 distinct protein structures, taken from the Protein Data Bank. The decomposition obtained correlates well with existing PFAM sequence classifications, but has the advantage of spanning the full protein, with the potential for novel domains. The coarse-grained network formed by the communities can also be used as a proxy for protein topology at the single-chain level; we demonstrate that grouping these proteins by their coarse-grained network results in a functionally significant classification. }
{community detection; protein structure; biological networks; spatial networks.}

% \keys{community detection; protein structure; biological networks; spatial networks.}
\end{abstract}

\section{Introduction}

All proteins are formed of chains of covalently bonded amino acids (also known as residues). The pattern of non-covalent bonding between units of the chain is what causes the protein to fold into its compact native structure; specifying the sequence of amino acids in a protein is sufficient to uniquely determine its folded shape \cite{Anfinsen73}. This structure then allows the protein to carry out its designated role within the cell.

Solving a protein's structure is costly in time and effort, yet the number of solved structures is growing rapidly. Over 130\,000 protein structures are now publicly available in the Protein Data Bank (PDB) \cite{Rose16}, and the size of this dataset is growing exponentially \cite{Berman13}. A widely-researched option for extracting insight from this dataset involves the search for protein domains; functional or structural subunits of a protein structure. Finding domains that are conserved between proteins helps to elucidate the relationship between a protein's structure and its function in the cell, and to classify the proteins into a taxonomy based upon their common structural features. The first efforts to assign protein domains were based upon manual expert curation \cite{Murzin95}. In recent years, two alternative databases involving both manual curation and computational assignment have emerged as mainstays; the CATH \cite{Dawson17} and SCOPe \cite{Fox13} databases. These databases focus on the domain as a structurally conserved unit, rather than as a compact, globular substructure, and as such the SCOPe and CATH labellings of the protein do not span the complete structure. Another widely used tool is the PFAM database \cite{Finn16}, which uses hidden Markov models to discover conserved regions of protein sequence.

One plausible alternative definition of a domain is that of a community on a protein structure network. Protein structure networks have been widely used in which the protein's amino acids are taken as the nodes of the network, with a wide variety of approaches taken to generate the edges, often using proximity of the $C_{\alpha}$ atoms (the central atom in each amino acid, bonded both to the amino acid's side chain and to the neighbouring amino acids via peptide bonds) \cite{Yan14}. This abstraction has shown promise in analysing individual proteins to identify key residues (amino acids) in allosteric communication \cite{DelSol07, DiPaola15, Amor16, Amitai04} and protein  thermal stability \cite{Csermely13}.  Tools have been developed to assist with the creation and visualization of the networks \cite{Chakraborty16_2, Doncheva11}.

The community structure of protein structure networks has been previously studied for individual proteins \cite{Delvenne10, Delmotte11, Zhang17}, showing that the community structure often aligns well with intuitive functional domains. Other work \cite{Tasdighian13, Hleap13} has validated network-based clustering over more traditional spatial clustering methods such as k-means clustering \cite{Jain10} and average-linkage clustering \cite{Feldman12}.

However, previous network-based methods \cite{Tasdighian13, Hleap13} have yet to be scaled to the set of proteins as a whole, possibly due to the computational cost involved. In this work,  we provide a comparison of network communities to known domain assignments for a large set of distinct proteins (4000 non-redundant protein chains).  We offer an approach using the Infomap community detection method, which uses the compression of a random walker's movement on the network to detect hierarchical community structure \cite{Rosvall11}. This notion of hierarchical community structure is required in order to account for the known multi-scale structure of proteins. We introduce a modified Jaccard measure to validate the generated community structures, and investigate the coarse-grained networks obtained by condensing each community into a single node, as a proxy for protein architecture.

Non-network-based comprehensive studies of protein structure such as \cite{Feldman12} only compare the numbers of domains found, not the assignments of residue positions to domains. Such approaches would therefore also not allow us to generate condensed networks of modules, and ignore or discard information about the hierarchical nature of community structure, for example by choosing a single cut-off point for the clustering dendrogram \cite{Feldman12}.\vspace*{-2pt}

\section{Methods}
\label{sec:Methods}

The analysis consists of three steps: the generation of the network from the protein structure, the community detection on the network and the storage and analysis of the communities as regions of the protein.

\vspace*{-4pt}
\subsection{Network generation}

There are many plausible approaches to generating a network representation of a protein's structure. The nodes of the network could be either the protein's atoms \cite{Amor16} or residues \cite{Yan14}. For a residue network, the edges are generated if two residues are within a certain distance. This distance measure can be based upon the inter-$C_{\alpha}$ distance, the inter-$C_{\beta}$ distance, or on the number of pairs of atoms within a certain proximity. Previous literature \cite{Yan14,Chakraborty16_2} has established a cut-off distance of $8\,\AA$ for $C_{\alpha}$ or $C_{\beta}$ networks, and $\sim 5\,\AA$ for networks based on the number of neighbouring atoms.

Here, a na\"{\i}ve yet flexible approach to network generation is used, which can generate either atomic networks or residue networks as required. Given the atomic positions from a PDB file, we let the atoms be nodes of the network. Undirected edges are then generated between atoms that are closer than a given cut-off distance. The cut-off distance between atoms $i$ and $j$ is defined as $c_{ij} = s \left( r_{i} + r_{j} \right)$, where $r_i$ is the covalent radius of atom $i$, and $s$ is a scaling parameter that can be varied to generate a network with higher or lower connectivity as required. If an atomic network is required, the edges are linearly weighted by proximity of the relevant atoms. If a residue network is required, the network is condensed by letting the amino acids be nodes in the network, with edges weighted according to the number of neighbouring atoms in the original atomic network. In what follows, residue networks with a value of $s=4$ are used, following \cite{Yan14}.

Performing this analysis on a protein with multiple chains often results in a network with distinct connected components, corresponding to each chain. As such, for this analysis the proteins are first split by chain. This helps ensure that any results are fixed at the sub-quaternary level.

Using a network generation tool written in Rust \cite{Matsakis14}, PDB files containing 10\,000 atoms can be parsed in this way in under 1 s.

\subsection{Community detection}

In choosing a community detection algorithm, we require a method that does not require the length scale or number of communities to be specified beforehand; we also require a method that is fast enough to allow for all 130\,000 proteins in the PDB to be analysed in a reasonable timeframe. We need the method to detect hierarchical community structure, in order to investigate the multi-scale structure of the protein, and a method with a resolution limit that will not impede the discovery of domain-level structure. Infomap \cite{Rosvall11} satisfies these constraints, along with known accuracy on benchmark graphs. Infomap has the disadvantage that it is prone to overpartitioning networks with geometric constraints, including spatial networks such as those generated in this work \cite{Schaub12}. However, empirically we see that the partitions generated correspond well to the domain-level structure of the protein (see overleaf).

\subsection{Storage}

All networks, partitions and results are stored in a MongoDB database \cite{Chodorow13}. This prevents duplication of effort; for a given parameter set, the database is first queried for the relevant information. If not found, then the relevant calculation is performed and the results stored in the database. In this way a large data set of protein structures with their community structure can be acquired.

\subsection{Performance}

In order to compare the match between the structure found using community detection and that found using other methods, we need a quantitative measure of similarity \cite{Fortunato16}. Traditional performance metrics such as the Normalized Mutual Information are unsuitable for this task; the predicted structure (for instance the PFAM domain structure \cite{Finn16}) generally occupies only a subset of the protein, whilst the generated community structure tiles the protein completely. Extra structure outside the region spanned by the prediction should not be penalized.

To this end, we modify the Jaccard index (JI), as follows. The JI is defined as the intersection between two sets, divided by their union, where in this case the sets correspond to regions of the protein sequence. This index is modified as follows:

For each `expected' domain:
\begin{itemize}
  \item Calculate the JI for all generated communities that overlap with the expected domain, i.e. $\frac{A \cap B}{A \cup B}$, where $A \cap B$ is the size of the overlap and $A \cup B$ the total length of sequence spanned by either the expected domain or the generated community.
  \item Perform an average of all the calculated JIs, weighted by the proportion of the total expected domain spanned by each community.
\end{itemize}

\begin{figure}[h]
    \begin{center}
        \includegraphics[width=0.95\textwidth]{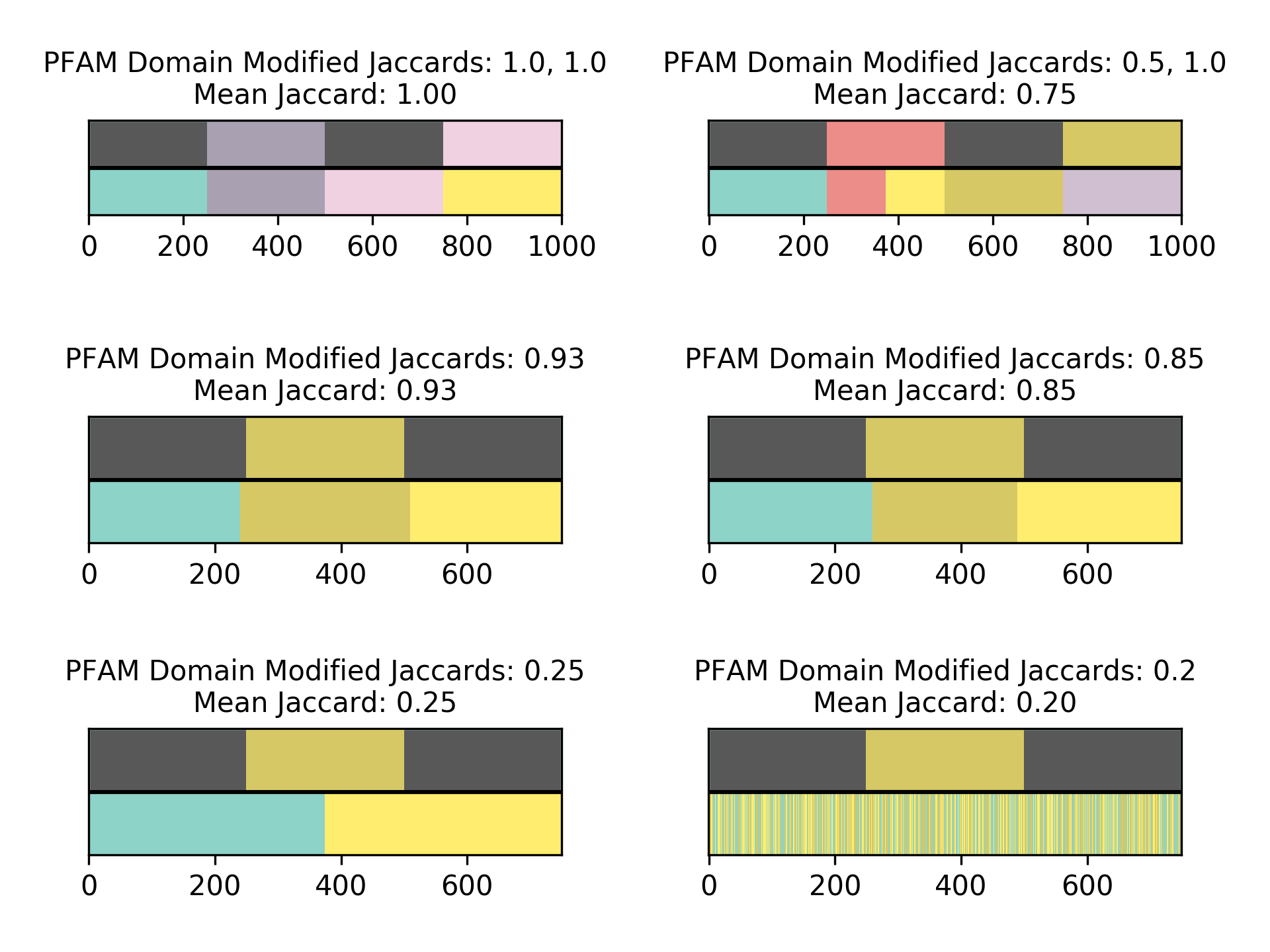}
          \caption[Modified Jaccard Test]{An illustration of the modified JI on example data. The expected domains (for instance, PFAM domains) are given above, and example community structures below for six possible cases. Each coloured block indicates a domain, with grey indicating unannotated regions. The upper right example shows a perfect score, as each PFAM domain matches a community perfectly. In contrast, the upper left figure shows that one PFAM domain has been split into two communities, giving the matching to that domain a score of 0.5 and an average score for the total match of 0.75. Note that the lower right figure, representing roughly the poorest imaginable case (a randomly shuffled two-community partition), still achieves a modified JI of 0.2.}

  \label{fig:modifiedJaccardTest}
    \end{center}
\end{figure}

This gives a score for each expected domain in the protein, indicating how well it is reflected in the community structure. On test data, this modified JI performs reasonably (see Fig.~\ref{fig:modifiedJaccardTest}), giving high scores to close matches and low scores to poor matches. Note that like the original JI \cite{Fortunato16}, this score does not take values in the full range $[0,1]$.

\newpage
In order to calculate the significance of a given modified JI, we use the \textit{z}-score. This is defined as:
\begin{equation*}
  z = \frac{\tilde{J} - \mu}{\sigma}
\end{equation*}

Where $\tilde{J}$ is the modified JI between the expected and generated partitions. $\mu$ and $\sigma$ are the average and standard deviation of the modified JI between the expected partition and a set of null models. $\mu$ therefore indicates the modified JI expected by chance. A \textit{z}-score of two indicates that the modified JI between the generated and expected partitions is two standard deviations higher than the expected value, and therefore, corresponds to a \textit{p}-value of $\sim0.02$ (assuming a normal distribution).

\begin{figure}
    \begin{center}
        \subfloat[][1CKI]{\includegraphics[width=0.22\textwidth]{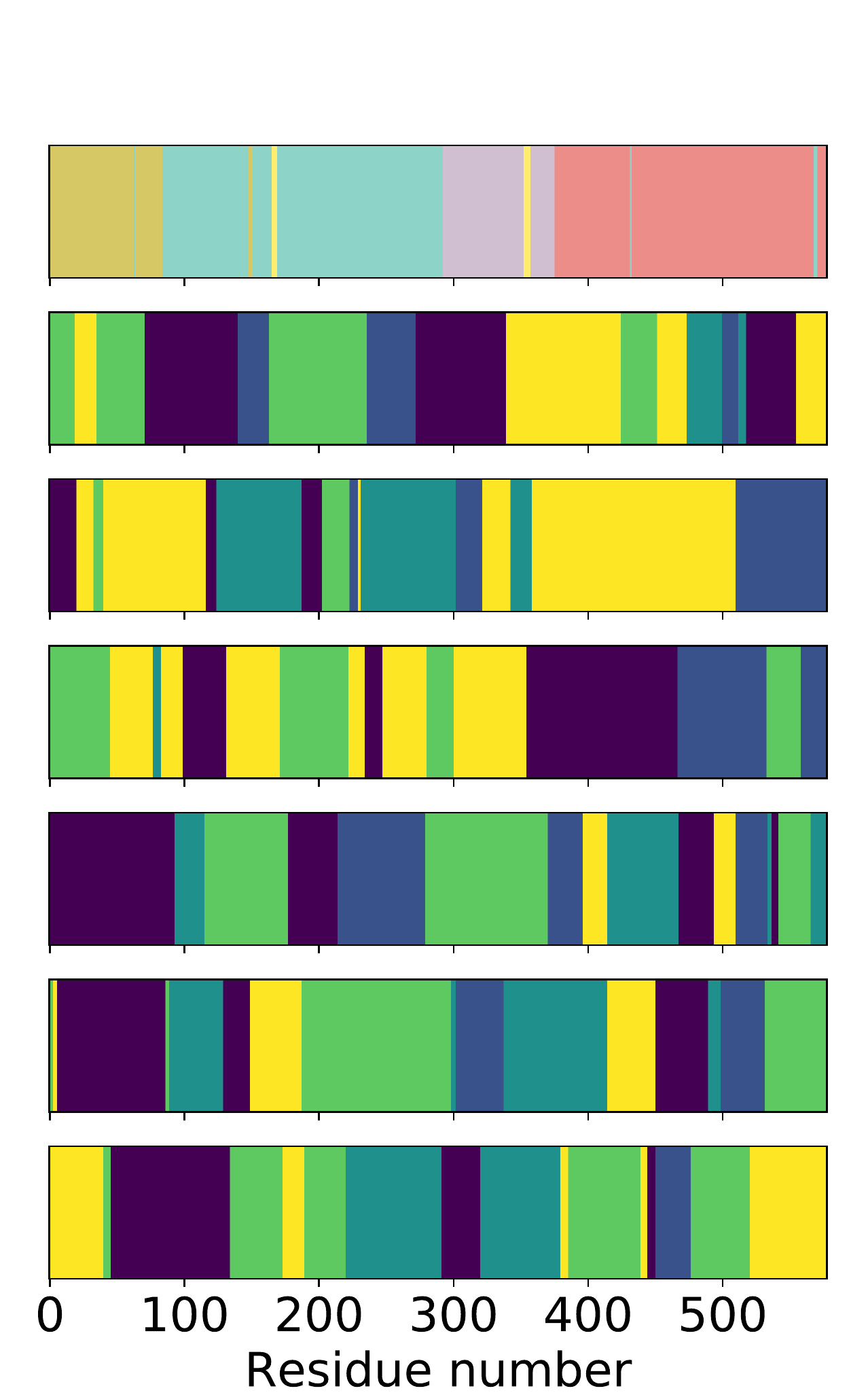}}\quad
        \subfloat[][1H3Q]{\includegraphics[width=0.22\textwidth]{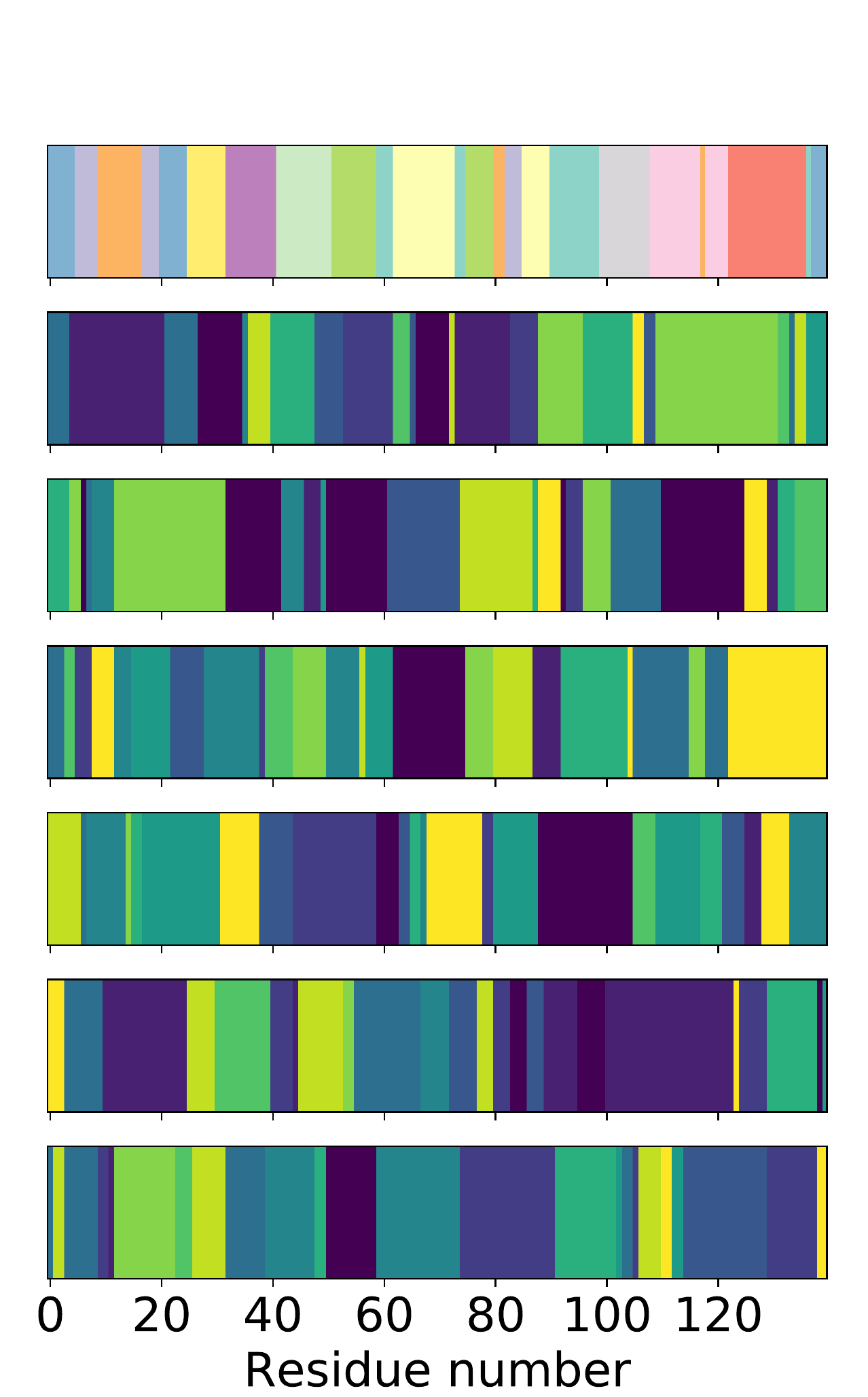}}\quad
        \subfloat[][3LMZ]{\includegraphics[width=0.22\textwidth]{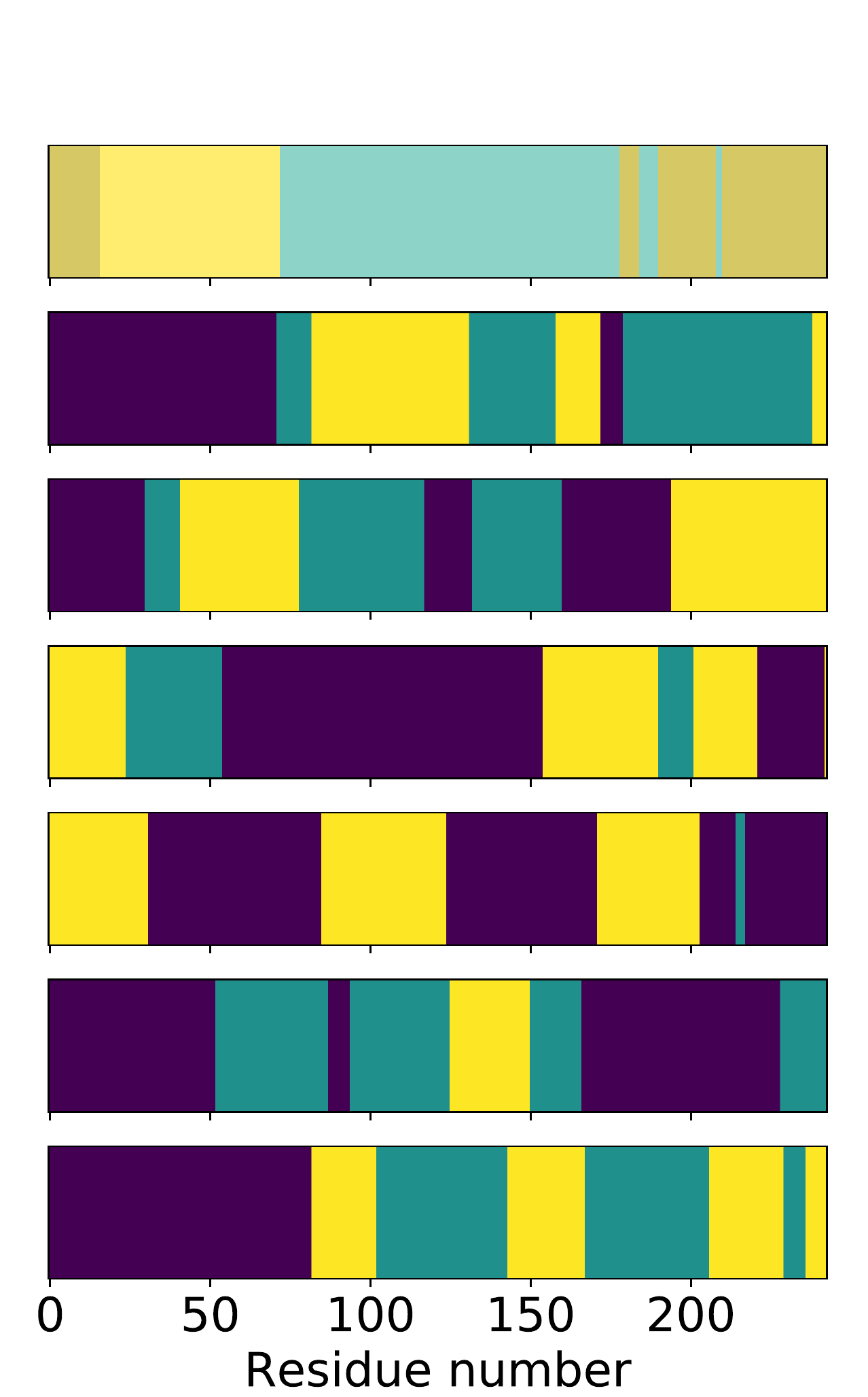}}\quad
        \subfloat[][2QTI]{\includegraphics[width=0.22\textwidth]{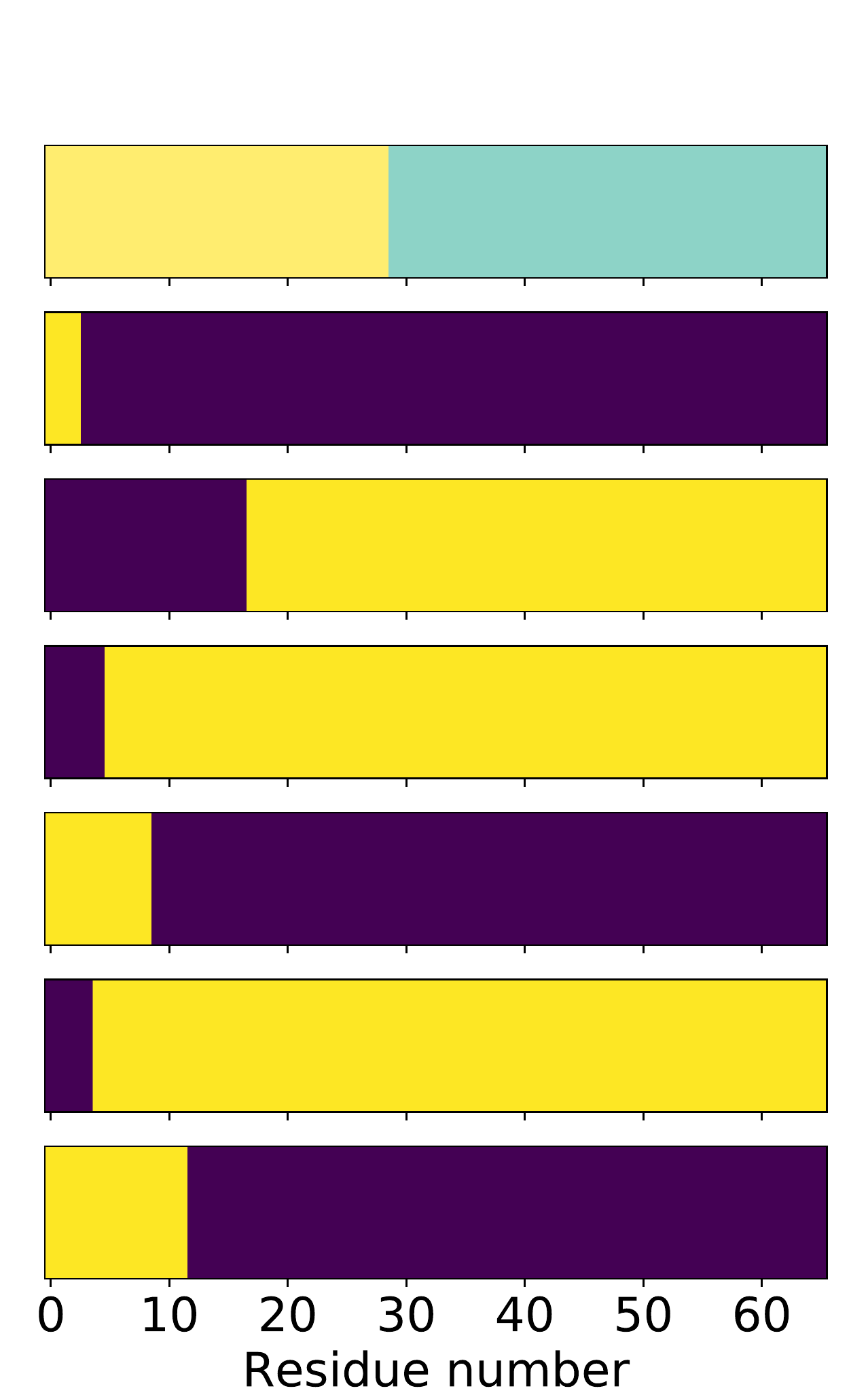}}

        \caption[Randomly generated null models]{Randomly generated null models, such that the number of boundaries between communities, and the total number of communities, is preserved. The generated structure to be tested is shown above, with six null models shown below. These null models succeed in capturing the properties of the test structure, ensuring that the comparison between null model and test is fair. }
        \label{fig:nullmodels}
    \end{center}

\end{figure}

These null models should be randomly generated, sharing some key properties of the generated community structure. In this work, the null models are created by constraining the number of boundaries (changes from one community to another along the sequence), and the total number of communities. Boundaries and community labels are then placed randomly to obey these constraints. Figure \ref{fig:nullmodels} shows the community structure to be tested above, with six generated null models below. These models succeed in capturing the rough features of the generated structure, whilst preserving randomness.

\section{Results}
\label{sec:Results}

% The results shown in what follows will all be single-chain amino acid networks, taken from a non-redundant subset of the Protein Data Bank (i.e. one without duplicate or near-duplicate structures), with threshold-scaling parameter $s=4$.

\begin{figure}[h]
    \begin{center}
% \vspace{-1cm}
        \subfloat[][Generated]{\includegraphics[width=0.3\textwidth]{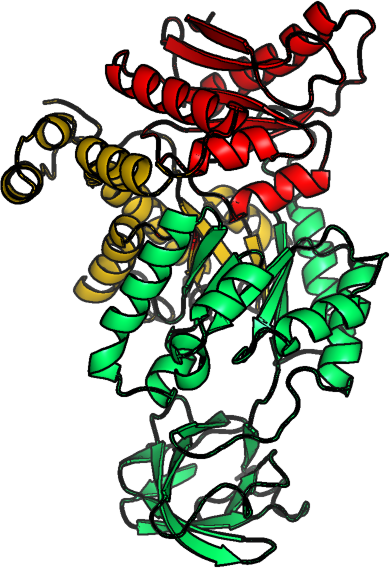}}\quad
        \subfloat[][Expected]{\includegraphics[width=0.3\textwidth]{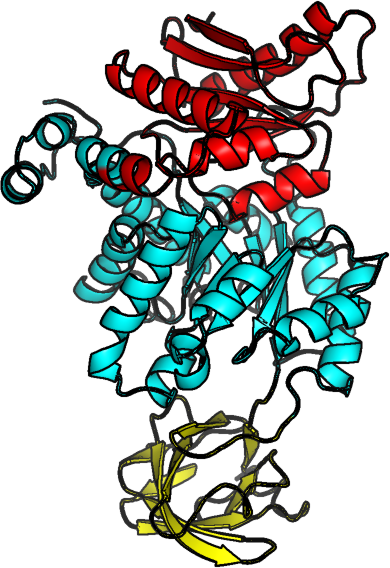}}

        \caption[1PKN]{The generated community structure (a) using a residue network with scaling parameter $s=4$ compared  with the known SCOPe domains (b) for a pyruvate kinase with PDB code 1PKN. One colour signifies one domain/community. Here, we see that one of the communities matches well to the existing SCOPe domain (both shown in red). }
        \label{fig:1pkn}
    \end{center}
    \vspace{-2cm}

\end{figure}

\begin{figure}[h]
\centering
\includegraphics[width=0.9\textwidth]{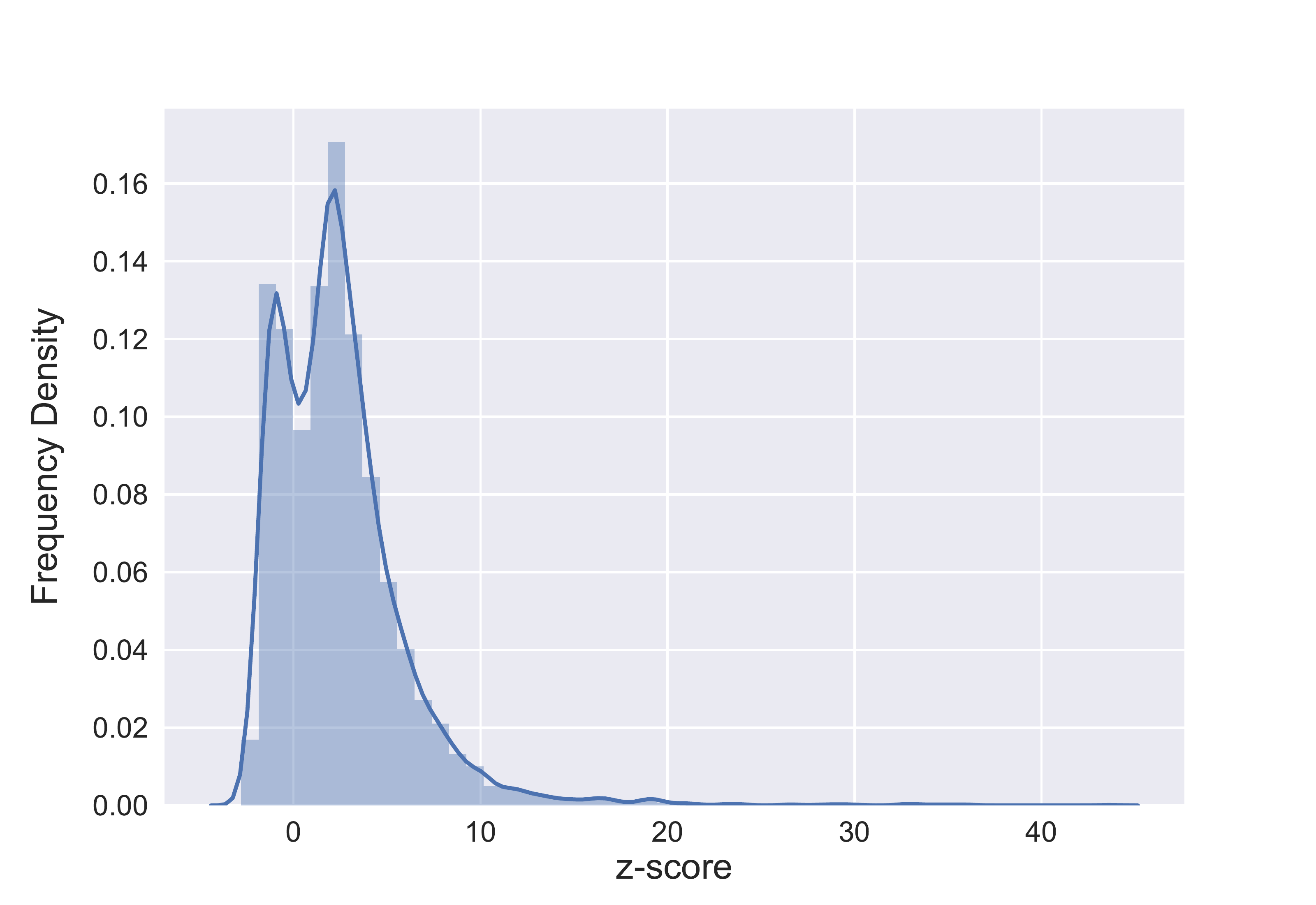}
    \caption{Histogram showing the \textit{z}-score for the modified JI between the generated community structure, and the PFAM domains, for $\sim$1000 test proteins, showing that in many cases the agreement is extremely significant.}
    \label{fig:zscores}
\end{figure}

Empirically, we see that a scaling parameter of approximately 4 gives communities corresponding to compact, globular regions of the protein structure (Fig. \ref{fig:1pkn}). We can quantify the extent to which these communities overlap with known protein annotations using the \textit{z}-score as defined previously. Here, we test the correspondence between the known PFAM domains, and the generated community structure.  In general, there is significant agreement, with the majority of proteins having a \textit{z}-score greater than 2 (Fig. \ref{fig:zscores}).
% Results on test proteins indicate that this scaling parameter gives globular, compact communities, corresponding well to known protein domains (\flag{Fig}.~\ref{fig:1pkn}).

\newpage

The communities found are based purely on the protein's structure, whilst the PFAM domains are based purely on sequence. As such, we expect discrepancies when the PFAM sequence domains correspond to more spatially extended, less well-connected regions of the structure. We can measure this by calculating the conductance of the regions of the network responding to the PFAM domains. If the set of nodes of a network $V$ is split into two subsets $S$ and $\bar{S}$, the conductance is defined as:
\begin{equation*}
    \phi(S) = \frac{\sum_{i \in S, j \in \bar{S}} A_{ij} }{\min(A(S), A(\bar(S))}
\end{equation*}

\begin{figure}[h!]
    \centering
    \vspace{-2cm}
    \includegraphics[width=0.85\textwidth]{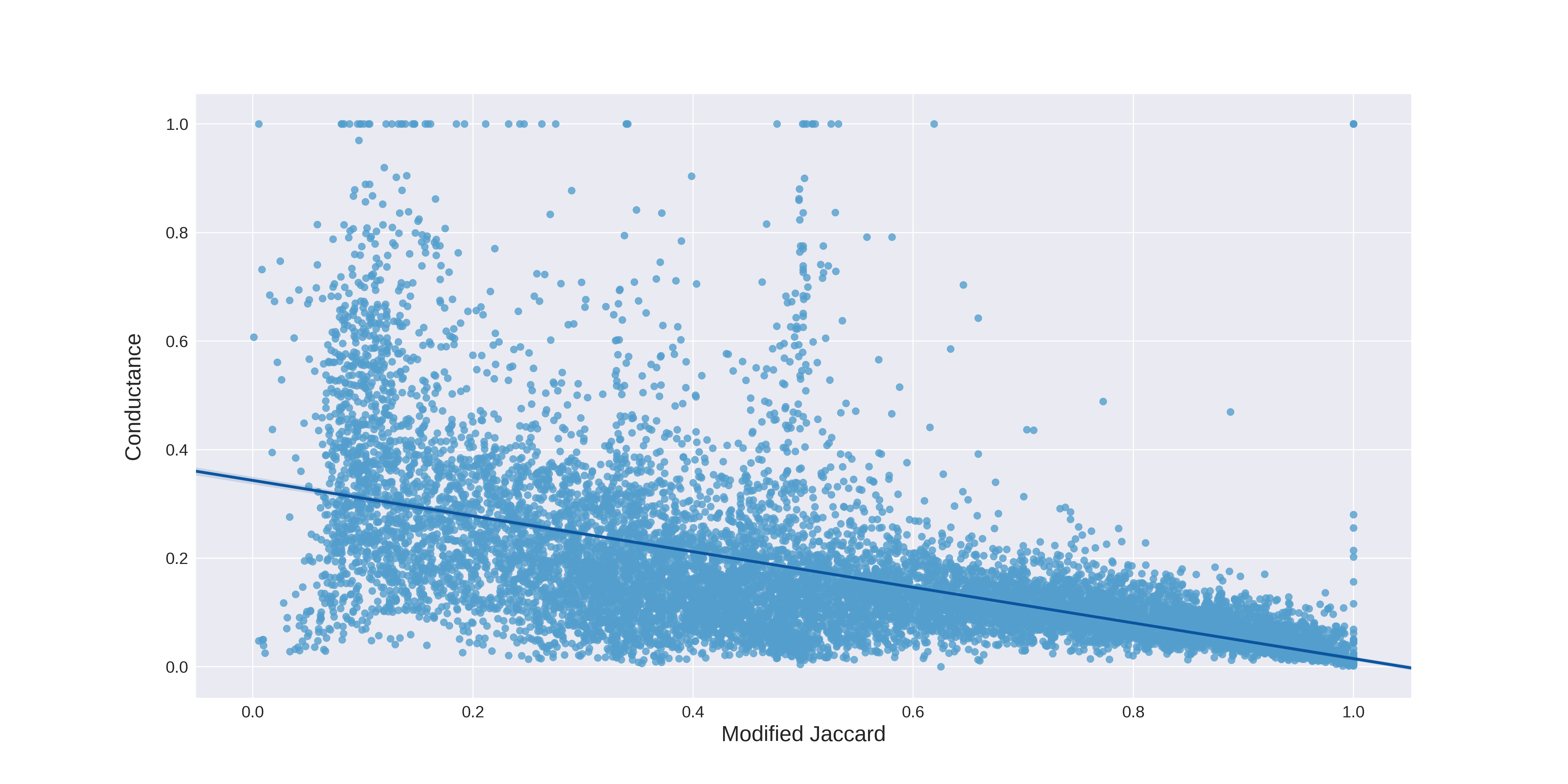}
    \caption{The conductance of the PFAM domain, when mapped onto the network, against the modified JI (indicating how well it corresponds to the community structure) The conductance is 0 for perfectly-isolated communities, and 1 for communities fully connected to the rest of the network, so we expect a negative correlation between modified JI and conductance; this is seen for the proteins studied here.} \label{fig:conductance}
% \end{figure}

% \begin{figure}[hb]
    \begin{center}
% \vspace{-10cm}
        \subfloat[][Modularity+Correlation]{\includegraphics[width=0.25\textwidth]{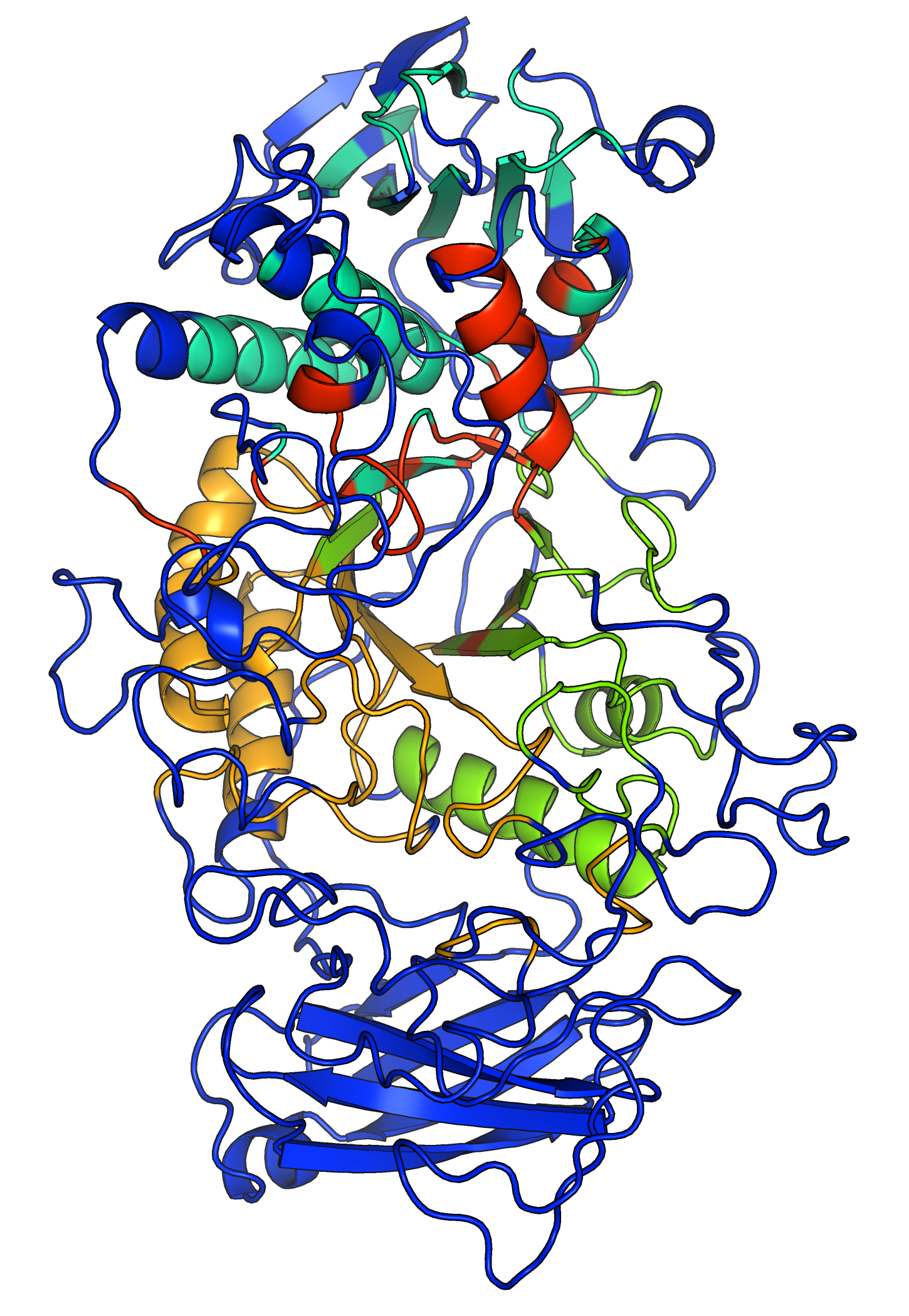}}\quad
        \subfloat[][Infomap]{\includegraphics[width=0.25\textwidth]{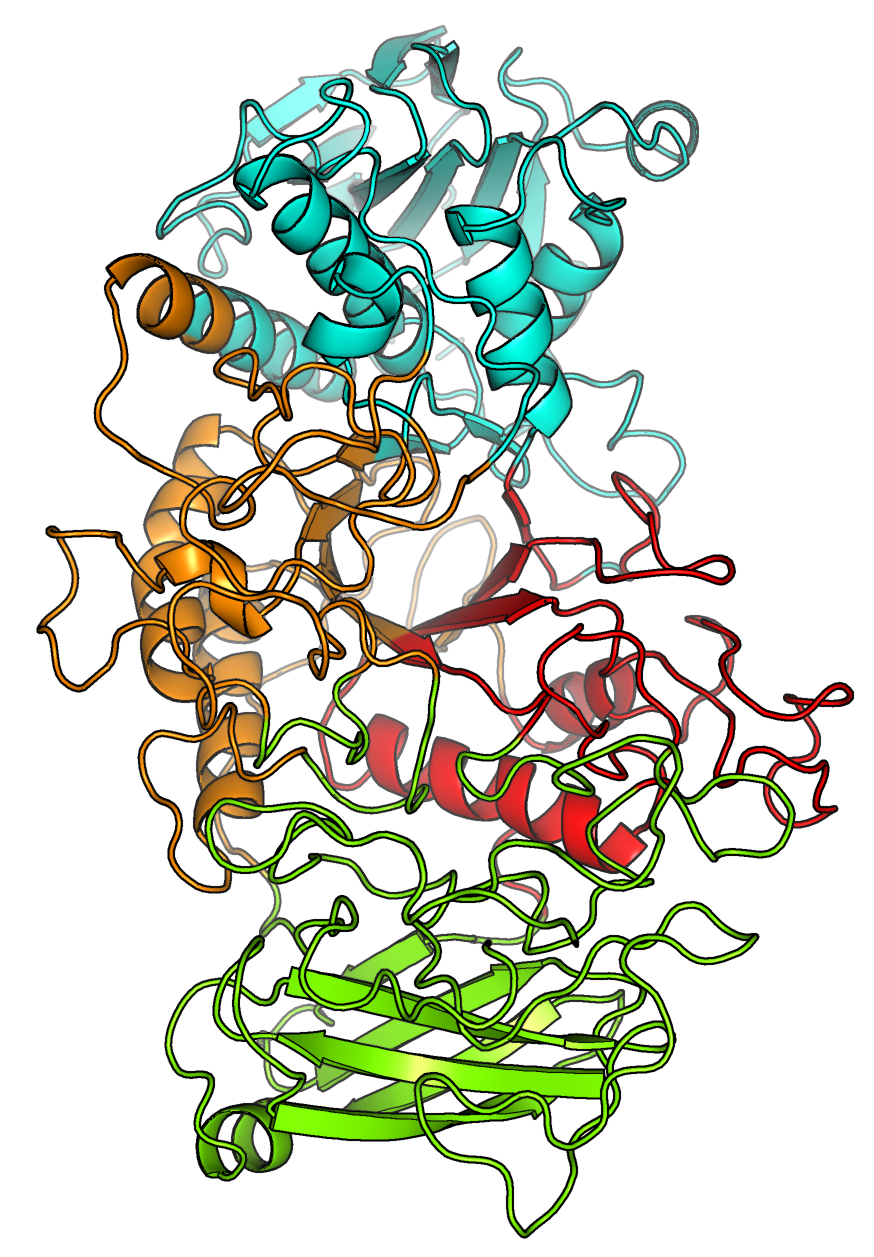}}\quad
        \subfloat[][SCOPe]{\includegraphics[width=0.25\textwidth]{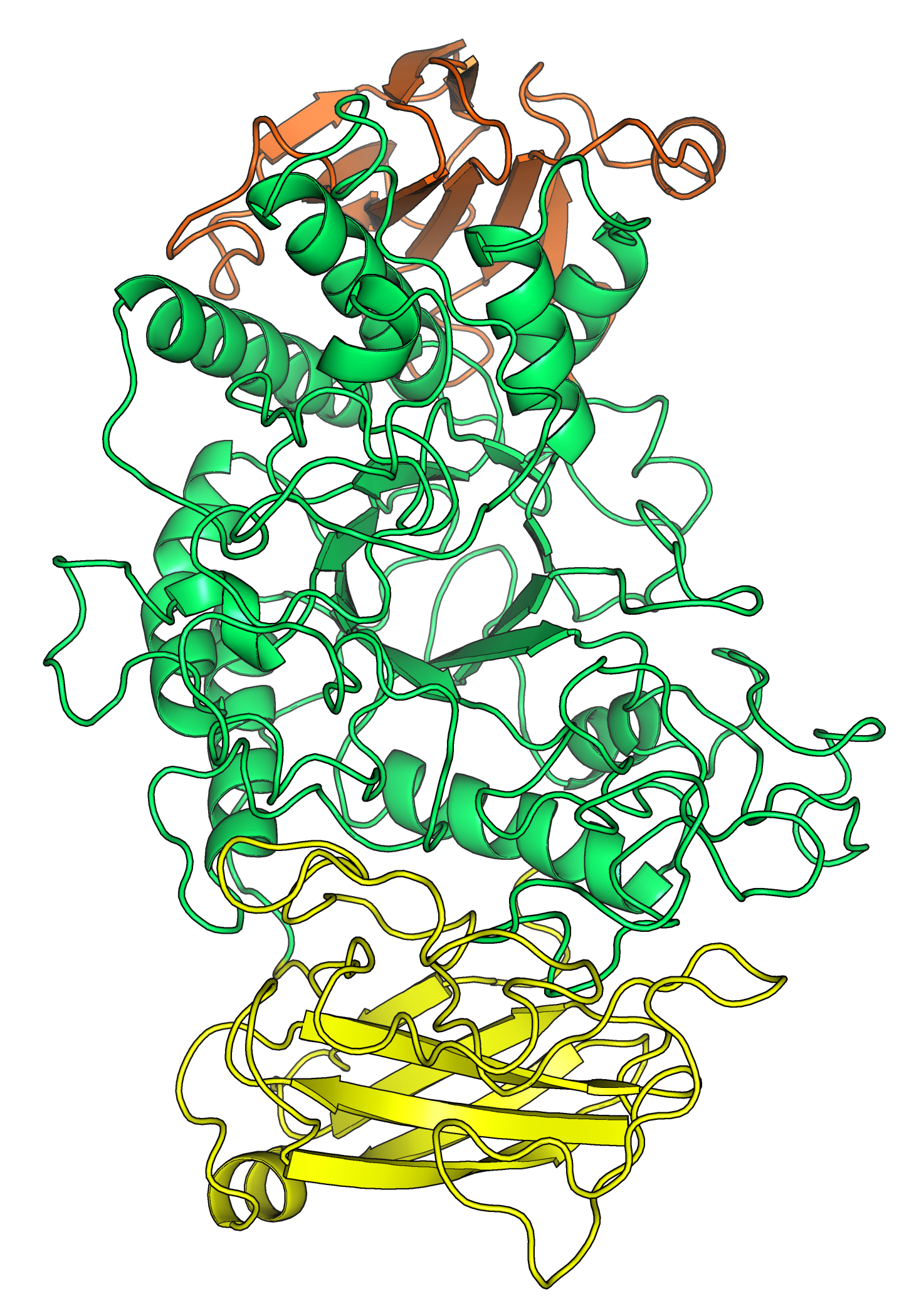}}\\
        \subfloat[][Sequence Labelling]{\includegraphics[width=0.7\textwidth]{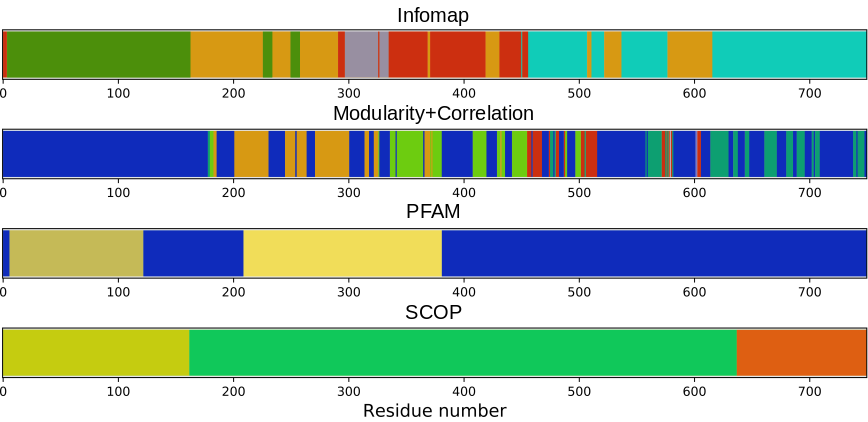}}
% \centerline{\includegraphics{OP-COMN180014f6.eps}}
        \caption[1BF2]{A comparison of annotations of the protein 1BF2. (a) The decomposition generated in previous work using a modularity-based method, combined with residue correlation analysis \cite{Hleap13}. Dark blue regions correspond to unannotated regions of the structure. (b) The decomposition using Infomap presented in this article. (c) The domains listed in the SCOPe structural domain database. (d) The same comparison, along with the PFAM annotations, presented as labellings of the protein sequence. Again, dark blue represents unannotated regions of the sequence. }
        \label{fig:1bf2}
    \end{center}
    \vspace{-2cm}
\end{figure}

Where $A_{ij}$ are the elements of the networks adjacency matrix, and $A(S) = \sum_{i \in S}\sum_{j \in V} A_{ij}$. Hence $\phi(S) \in [0,1]$, with a lower conductance corresponding to a more isolated region of the network. We expect the modified JI and the conductance to negatively correlate; Figure \ref{fig:conductance} shows this is indeed the case.

\begin{figure}[h]
\centering
\includegraphics[width=0.9\textwidth]{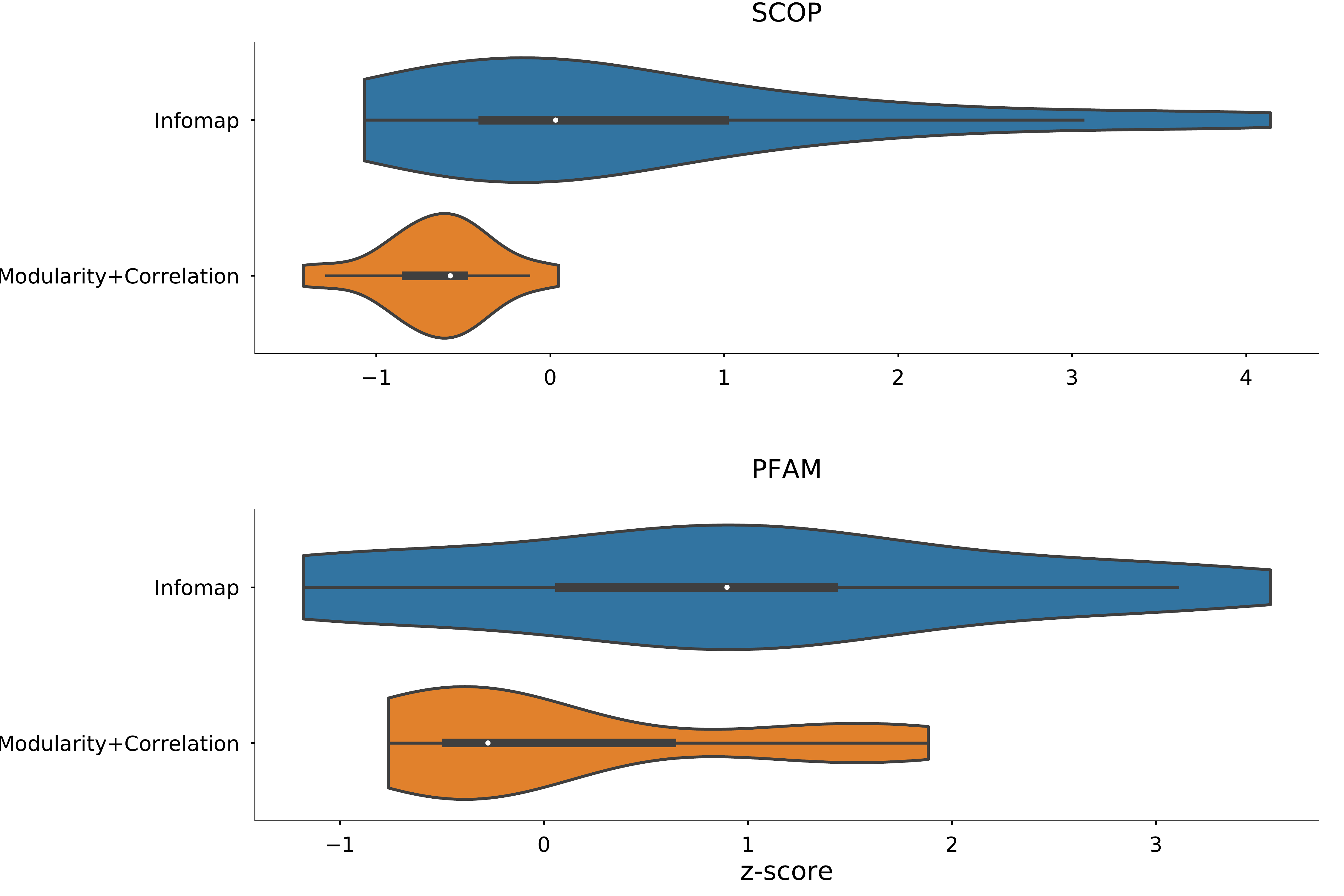}
        \caption[zScoreComparison]{A comparison of the \textit{z}-scores for a set of 20 reference proteins, giving the significance of the overlap between SCOP (above) and PFAM (below) for the modularity + correlation method \cite{Hleap13} and the proposed Infomap-based method. We see that in both cases the Infomap-based method has a more significant similarity to existing annotations. }
        \label{fig:fullStats}
\vspace*{10pt}
\end{figure}

We can compare the communities generated using Infomap to previous network-based attempts to assign domains, which used correlation networks and a modularity-based method \cite{Hleap13}. Figure \ref{fig:1bf2} compares these results qualitatively to SCOPe annotations and to the results obtained using our protocol. Figure \ref{fig:fullStats} compares the results quantitatively, using the \textit{z}-score. A drawback of the correlation-based approach is that a set of homologous proteins is needed; our method has the advantage that it can be performed on single proteins, meaning that the partition spans the full protein structure, and making the approach scalable to larger datasets.

\newpage
\newpage

In addition to the communities' potential value as structural domains, the arrangement of the communities may be used as a proxy for topology. The community structure can be converted to a coarse-grained network in which the protein's communities become nodes, linked if the respective communities are neighbours. We can then classify the proteins according to the arrangement of their communities, by grouping proteins with isomorphic coarse-grained networks.

If the community structure is truly capturing the protein's topology, we expect this grouping to reveal aspects of protein function. We can test this claim using Gene Ontology (GO) term analysis  \cite{Ashburner00}. This effort assigns functional relevance (e.g. lactase activity, oxidoreduction) to genes. The SIFTS project \cite{Velankar13} maps these GO terms to records in the PDB, meaning that each protein now has a set of labels encoding information about its function in the cell. We can then test if the grouping results in enriched GO terms, i.e. terms appearing more often than expected by chance \cite{Huang08,Rhee08}. For $N$ total proteins, and a subset of that dataset with $n$ proteins, the probability of a GO term being found is given by the cumulative distribution function (CDF) of the hypergeometric function. For a given GO term, let $k$ be the number of times it occurs in the subset, and $K$ be the number of times it occurs across the full dataset. Then the likelihood that the term would be seen $k$ times by chance is:
\begin{equation*}
  CDF = 1 - \frac{{n \choose k+1} {N - n \choose K - (k+1)} }{ {N \choose K} } \; {}_3 F_2 \left[ \begin{matrix}1, \: \: \: & &k+1-K,& &k+1-n\\ &k+2,&&N+k+2-K-n&\end{matrix} \; \; ;1 \right]
\end{equation*}

Where $_3F_2$ is the generalized hypergeometric function. From the CDF, we can acquire p-values for a given grouping and GO term; we consider GO-terms with a \textit{p}-value of less than 0.01 to be enriched in the subset to a statistically significant extent. As we testing $M$ distinct GO terms, we account for multiple hypothesis testing by applying the Bonferroni correction. If comparisons of $M$ GO terms are being made, the raw \textit{p}-value is multiplied by $M$ to give a more conservative estimate of the likelihood.

\begin{figure}[hb!]
\centerline{\includegraphics[width=0.4\textwidth]{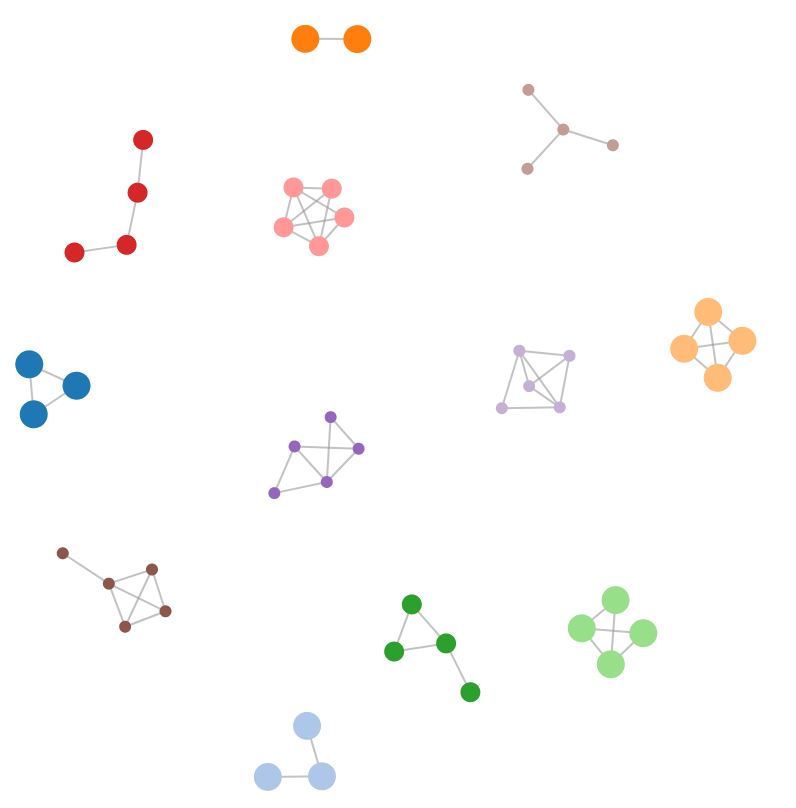}}
    \caption{The coarse-grained networks generated from the community structures of approximately 4300 protein chains, taken randomly from a non-redundant subset of the Protein Data Bank. Only coarse-grained networks common to at least three proteins are shown (accounting for $\sim 3900$ proteins in total). The node size is proportional to the number of proteins exhibiting that coarse-grained network. } \label{fig:singleChainTopology}
% \vspace*{10pt}
% \end{figure}

\vspace{3em}
% \begin{table}
    % \processtable{{The ten most common protein topologies in the dataset studied, ordered by prevalance}\label{tab:singleChainTopology}}
    \begin{tabular*}{\textwidth}{@{\extracolsep{\fill}}ccc} % {\textwidth}{ccc}  %{@{\extracolsep{\fill}}ccc}
    % \toprule
            Coarse-grained network & Number of enriched GO terms (\textit{p}\textless  0.01) & Number of proteins\\
    \hline\\
    \parbox[c]{1em}{\includegraphics[height=0.7cm]{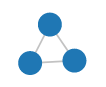} } & 331 & 1725 \\
    \parbox[c]{1em}{\includegraphics[height=0.7cm]{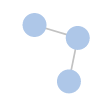}} & 130 & 307 \\
    \parbox[c]{1em}{\includegraphics[height=0.7cm]{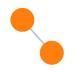}}& 116 & 841 \\
    \parbox[c]{1em}{\includegraphics[height=0.7cm]{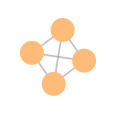}}& 104 & 445 \\
    \parbox[c]{1em}{\includegraphics[height=0.7cm]{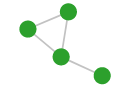}}& 34 & 55 \\
    \parbox[c]{1em}{\includegraphics[height=0.7cm]{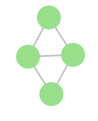}}& 52 & 207 \\
    \parbox[c]{1em}{\includegraphics[height=0.7cm]{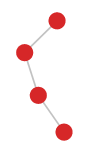}}& 48 & 26 \\
    \parbox[c]{1em}{\includegraphics[height=0.7cm]{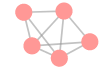}}& 23 & 19 \\
    \parbox[c]{1em}{\includegraphics[height=0.7cm]{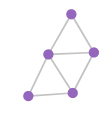}}& 9 & 6 \\
    \parbox[c]{1em}{\includegraphics[height=0.7cm]{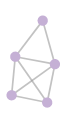}}& 22 & 8 \\
    % \parbox[c]{1em}{\includegraphics[height=0.7cm]{Figs/topology11.png}}&7 & 4 \\
    % \bottomrule
    \end{tabular*} %}{}
    % \vspace*{-9pt}
    \captionof{table}{The ten most common protein topologies in the data set studied, ordered by prevalance.}
    \label{tab:singleChainTopology}
% \end{table}
\end{figure}

We see that 90\% of the proteins studied can be represented by only 10 coarse-grained networks, all of which are associated with GO-term enrichment (Fig. \ref{fig:singleChainTopology} and Table \ref{tab:singleChainTopology}).

\newpage

\section{Conclusion}

There have been many attempts to define the domain, as a compact, repeated unit of protein structure. But choosing these compact, globular substructures in an automated way has traditionally been challenging. We present results showing that a simple weighted network of residue contacts analysed with Infomap can successfully fragment a protein into compact modules. By using a modified JI, we show that in general these modules correlate well with existing PFAM annotations, yet have the advantage that they span the full protein structure. This has potential applications in molecular dynamics and electron\break microscopy.

We also show that by generating a coarse-grained network, in which the communities of the network are taken as nodes, we can group a large set of proteins in a way that gives significant functional enrichment, as measured by the prevalence of GO terms. This suggests that the community structure can be used as a proxy for the protein topology.

The next step will be to use this approach to search for repeated communities with similar internal topology that have not yet been identified as domains, with the hope of establishing a new framework for domain discovery. %\vadjust{\vspace*{35pt}\pagebreak}

\section*{Funding}

This work was supported by the Engineering and Physical Sciences Research Council
Centre for Doctoral Training in Computational Methods Materials Science
[EP/L015552/1] to W.P.G. and the Royal Society and the Gatsby Foundation to
S.E.A.

\vspace{3em}
% \newpage

\end{document}